\documentclass[aps,prm,twocolumn,showkeys,superscriptaddress,unsortedaddress]{revtex4-2}

\usepackage{float}
\usepackage{mathtools}
\usepackage{color}
\usepackage{amssymb}
\usepackage{amsmath}
\usepackage{bm}
\usepackage{varioref}

\begin{document}

\bibliographystyle{apsrev}
\title{Structures and velocities of noisy ferroelectric domain walls}
\author{Nora Bauer}
\email{nbauer1@vols.utk.edu }
\affiliation{Department of Physics \& Astronomy, University of Tennessee, Knoxville, Tennessee 37996, USA}
\author{Sabine M. Neumayer}
\email{neumayersm@ornl.gov}
\author{Petro Maksymovych}
\email{maksymovychp@ornl.gov}
\affiliation{Center for Nanophase Materials Sciences, Oak Ridge National Laboratory, Oak Ridge, Tennessee 37831, USA}
\author{Maxim O. Lavrentovich}
\email{lavrentm@gmail.com}
\affiliation{Department of Physics \& Astronomy, University of Tennessee, Knoxville, Tennessee 37996, USA}
\begin{abstract}
Ferroelectric domain wall motion is fundamental to the switching properties of ferroelectric devices and is influenced by a wide range of factors including spatial disorder within the material and thermal noise. We build a Landau-Ginzburg-Devonshire (LGD) model of 180${}^{\circ}$ ferroelectric domain wall motion that explicitly takes into account the presence of both spatial and temporal disorder. We demonstrate both creep flow and linear flow regimes of the domain wall dynamics by solving the LGD equations in a Galilean frame moving with the wall velocity $v$.   Thermal noise plays a key role in the wall depinning process at small fields $E$. We study the scaling of the velocity $v$ with the applied DC electric field $E$ and show that noise strongly affects domain wall velocities. We also show that the domain wall widens significantly in the presence of thermal noise, especially as the material temperature $T$ approaches the critical temperature $T_c$. These calculations therefore point to the potential of noise and disorder to become control factors for the switching properties of ferroelectric materials, for example for advancement of microelectronic applications.
\end{abstract}

\date{\today}

\maketitle

\section{Introduction}

\begin{figure}[!ht]
\includegraphics[width=3.3in]{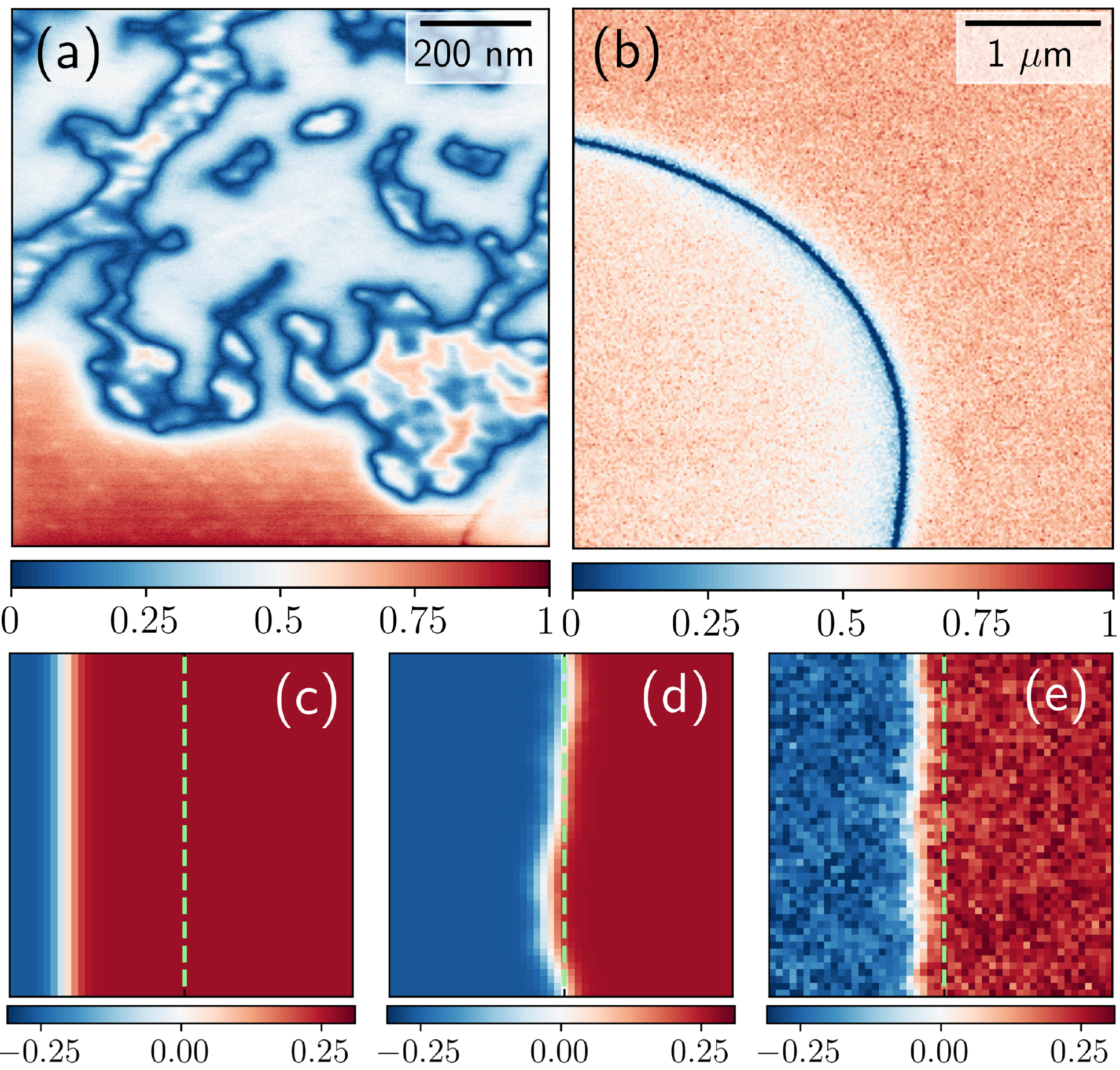}
\caption{\label{fig:domainwall}  (a) and (b) are Piezoresponse Force Microscopy images of domain walls in representative ferroelectric materials, CuInP${}_{2}$S${}_{6}$ (a) and Sn${}_{2}$P${}_{2}$S${}_{6}$ (b). The color indicates the response of the material to a locally applied AC electric field (via an AFM tip). We show our simulated domain wall under an applied field without noise in (c), with quenched spatial noise in (d), and with thermal and quenched spatial noise in (e). The light green dashed line indicates the initial position of the wall, which moves in response to an applied DC  electric field when the field (in combination with the thermal noise) is large enough to overcome the pinning due to the quenched spatial disorder, as in the case of (c) and (e). In (d), the domain wall is pinned by the quenched noise. The colorbars in (c-e) indicate the value of the polarization $P$ in C/m${}^2$.   }
\end{figure}

Domain walls have long been recognized \cite{merz1954} to play a critical role in the switching properties of ferroelectric materials and devices. Here we consider the motion of Ising-like, 180${}^{\circ}$ domain walls between two stable polarization domains of a ferroelectric thin film, using as a representative the Landau-Ginzburg potential of BaTiO${}_3$, barium titanate. Such walls are found in a variety of materials and it is possible to image them using piezoresponse force microscopy, as shown in Fig.~\ref{fig:domainwall}(a,b) for ferroelectric CuInP$_2$S$_6$ and Sn$_2$P$_2$S$_6$. It is apparent that domain walls [indicated by lines of low piezoresponse in Fig. 1(a,b)] can have different size and shapes. Moreover, in ferroelectric materials, both the thermal fluctuations and spatial disorder (in the form of defects, grain boundaries, dopants, etc.) play a key role in the mobility of the domain walls and, hence, the polarization switching behavior.

It has long been appreciated \cite{merz1954} that, at low applied electric fields $E$, the domain wall motion is governed by an activated process, strongly influenced by quenched disordered (defects) in the material. These processes result in creep motion \cite{rappe2007,ferrowalldepinning}, and a strongly non-linear dependence of the switching rate, or domain wall speed, on the applied field.  Various phenomenological explanations for this behavior, especially for 180${}^{\circ}$ domain walls in barium titanate, have been in the literature for  decades  \cite{weinreich}. Indeed, the dynamics of ferroelectric domain walls is a paradigmatic subject \cite{Tybell2002,morozovskawallwidth} and has been studied using a wide variety of methods due to its importance in most aspects of ferroelectric switching. Of direct relevance to the present paper is the transition between the domain wall creep and flow regime, which has been reported to occur in experiments \cite{ferrowalldepinning,ferrowallvelocity},  as well as in theoretical calculations \cite{rappe2007,rappe2016}. The primary driving force of such a transition in most previous works is the increasing strength of the field $E$, which depins domain walls above the so-called activation field threshold \cite{copolymerswitch,depolarization}. Here we explicitly bring out the effect of thermal noise and explore thermally-activated domain wall depinning, focusing on the noisy wall behavior in  ``multi-well''  ferroelectrics.

  The quantitative framework taking into account the interplay between the various sources of noise in these systems, including the spatially-quenched disorder (due to grain boundaries and defects) and thermal noise, is still lacking.  In this work we aim to systematically study, using a stochastic Landau-Ginzburg-Devonshire (LGD) free energy approach, the effects of quenched and thermal noise on the propagation and shape of the 180${}^{\circ}$ domain wall in a thin ferroelectric film under a uniform applied electric field $E$ (along one of the spontaneous polarization directions). The qualitative effects of these different noises are illustrated in Fig.~\ref{fig:domainwall}(c-e). Note especially that in the presence of spatial, quenched disorder, the domain wall might become \textit{pinned}, such that it is necessary to apply a large enough field  beyond a threshold ``depinning'' field $E_{dp}$, $E>E_{dp}$, in order to move the domain wall [see Fig.~\ref{fig:domainwall}(d)]. Several prior works also reported on LGD modeling in the presence of noise \cite{phasefield,phasefield2,Klotins2008}. Although computationally demanding, one of the primary motivations for including noise explicitly is the ability to model finite-temperature effects accurately. Here we extend these studies toward specific effects of noise on the structure and mobility of domain walls that underpin ferroelectric switching.

 We quantify the effect of the noise through a novel approach of monitoring the domain wall over an extended period of time by using a boosted frame of reference in which the domain wall remains stationary. In contrast to other approaches, such as the observation of the inflation of the periphery of small circular domains \cite{noisymagneticwalls,Tybell2002}, this technique allows for continuous observation of a domain wall of a fixed size. We find that, as in ferromagnetic systems, thermal noise mitigates pinning due to spatial disorder. In particular, significant acceleration of the domain wall velocity can be achieved with added noise. We also find that near the critical temperature (where the ferroelectric is described by a ``triple well'' potential), the thermal noise can significantly modify the structure of the domain wall, nucleating a large paraelectric region in the domain wall interior. Such domain wall broadening is associated with sharp increases in domain wall mobility, recapitulated in more detailed models which modulate the width ``by hand'' \cite{morozovskawallwidth}. The increase in mobility is due to the suppression of the critical depinning field $E_{dp}$ due to the presence of the paraelectric region within the domain wall, a markedly different behavior from the ferromagnetic case. Thus, we show that noise may play an important role in generating these wall shape changes in ferroelectrics, and, consequently, controlling the switching behavior.

Given the generality of the problem setting, the effects of the noise and disorder are anticipated to apply to many ferroelectric materials. Effective solutions of stochastic LGD equations can therefore extend decades of previous successes of this model in ferroelectric research, toward new approaches to designing new switching and domain wall devices, as well as new approaches to optimizing device energy efficiency. Results will be particularly significant in proximity to phase transitions, where the effect of thermal excitations are maximized.  The thermal excitations are relevant for electrocaloric applications \cite{electrocaloricreview}, in nanoscale ferroelectrics subject to confinement effects \cite{confinementreview}, and near the quantum critical regime \cite{quantumcritical1}.

The LGD approach considered here is of interest in a general context, as similar equations describe the growing edge of a cellular population on a surface (or a spreading epidemic) \cite{KorolevRMP}, a moving reaction front in a multi-component system \cite{reactionLGD}, the phase boundary motion in a binary metal alloy \cite{AllenCahn} (the classic application of such equations), and the domain wall motion in ferromagnets \cite{noisymagneticwalls}.  In most previous studies, including for the ferromagnets, the domain walls of interest connect two (meta)stable phases, separated by some potential barrier. In this case, it is often possible to model the domain wall as a fluctuating line or surface, with vanishing width \cite{LGwalls,sharpinterface}. In other words, the center of mass of the domain wall may be treated as an elastic line moving in a disordered medium (in the presence of quenched spatial disorder and thermal noise), which captures the essential physics of depinning and creep motion. In this work, we will study the profile of the domain wall in more detail. We shall see that the elastic line model may not always be suitable, especially near the phase transition to a paraelectric state, where the domain wall thickness can strongly vary with the strength of the noise.

\section{Methods}

\subsection{LGD Model}

We consider an Ising-like domain wall separating two stable polarization states pointing in opposite directions (a so-called 180${}^{\circ}$ wall). The polarization states  are described by a scalar order parameter $P(\mathbf{x},t)$ with values $\pm P_0$ corresponding to the two stable polarizations.  The free energy functional of the order parameter reads
\begin{equation}
\psi[P]=\frac{\kappa}{2} (\nabla P)^2+ V[P]-EP, \label{eq:LGDpotential}
\end{equation}
with $E$ the applied field, $V[P]$ the thermodynamic potential, and $\kappa$ the gradient energy coefficient. For the well-studied material barium titanate, the thermodynamic potential $V[P]$, along with the temperature dependence, is readily available in, e.g.,  Ref.~\cite{LGDcoeffs}.   For concreteness, we will use the potential described in this reference.  It is given by
\begin{equation}
V[P]=\alpha_1 P^2+\alpha_{2} P^4+\alpha_{3}P^6, \label{eq:potential}
\end{equation}
with the coefficients $\alpha_i$ tabulated in Table~\ref{tb:coefficients}.

 \begin{figure}[!ht]
\includegraphics[width=3.2in]{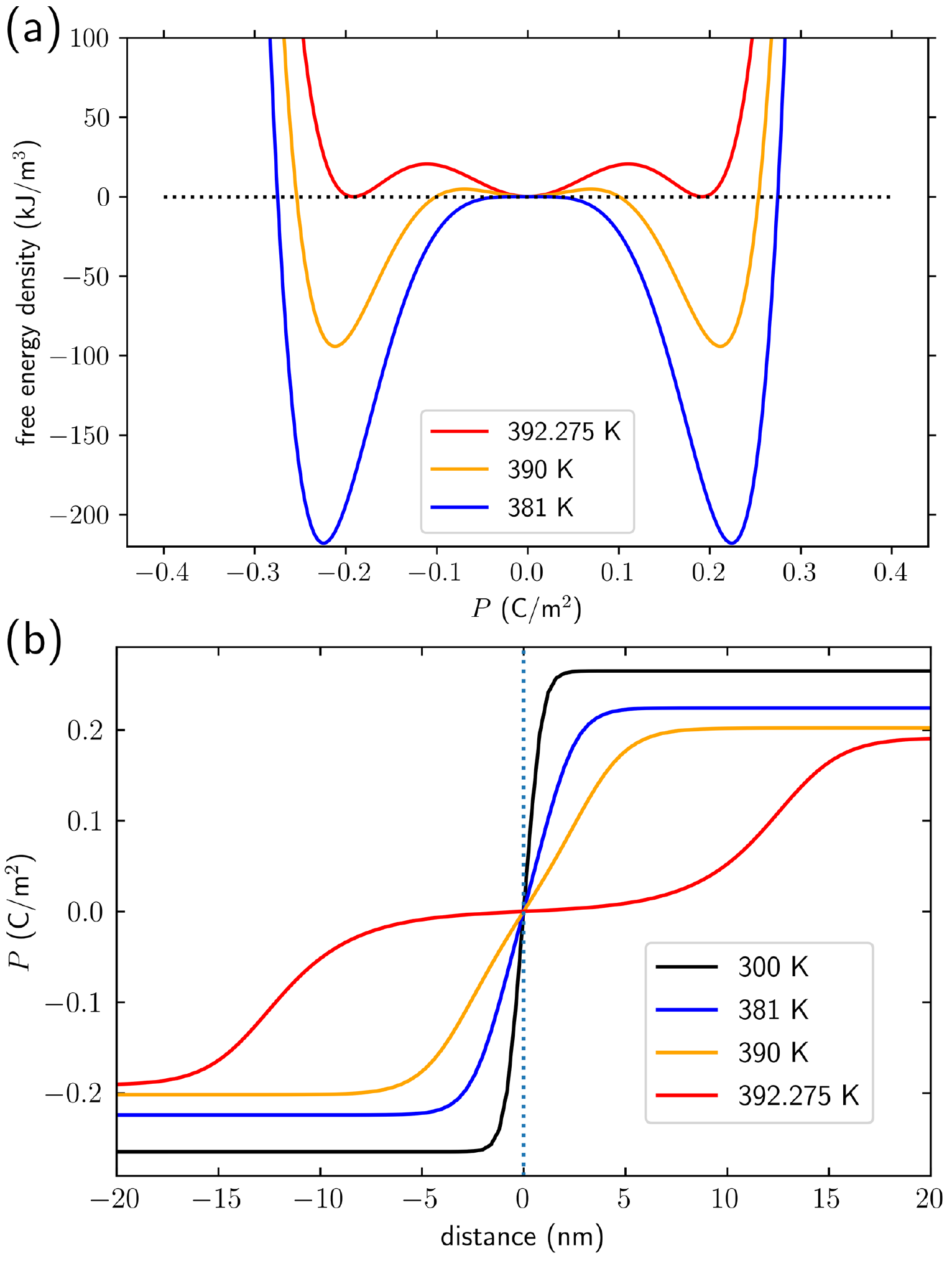}
\caption{\label{fig:potentialshapes}  (a) The LGD potential used in this study at various temperatures $T$. Note the development of the metastable paraelectric state (with polarization $P=0$) as we increase $T$ to right below the critical temperature $T_c =392.277\ldots$~K.  This metastable state has profound consequences for the profile of domain walls, shown at various temperatures (without noise) in (b), where the vertical dashed line indicates the center of the domain wall. As $T\ \rightarrow T_c$, the domain wall widens and eventually develops a plateau at $P=0$ (red line). These shapes were derived using the numerical solution to the LDG equation and are in perfect agreement with the analytic predictions (see Appendix).  }
\end{figure}

The coefficient $\kappa$ will generate an energy penalty to variations of the polarization $P$ and will result in an associated domain wall line tension.  The value of $\kappa$ will depend on the phase of the material, and may also generally be a tensorial quantity, as the domain wall tension is generically anisotropic \cite{LGDcoeffs2}. Here we will consider an isotropic value for simplicity. The coefficients $\alpha_i$ in Eq.~\eqref{eq:potential} depend on temperature $T$ and the potential $V[P]$ exhibits two minima at $P = \pm P_0$ for all temperatures $T<T_c = 392.2773660075\ldots$~K.  We find that, in terms of the coefficients of the potential,
\begin{equation}
P_0^2=- \frac{\alpha_2}{3\alpha_3}\left(1+\sqrt{1-\frac{3\alpha_1\alpha_3}{\alpha_2^2}} \right). \label{eq:P0}
\end{equation}
In addition, when $T^* \equiv 381~\mathrm{K}<T<T_c$, there is an additional local minimum at $P=0$, corresponding to a (metastable) paraelectric state.
 Shapes of the potential $V[P]$ at different temperatures are shown in Fig.~\ref{fig:potentialshapes}(a).
Note that at $T=T_c$, the local paraelectric minimum at $P=0$   becomes a global minimum, along with the other two minima at $P= \pm P_0$. As we shall see, this has important consequences for the shape of the domain wall and the behavior of the system in the presence of noise.

\begin{table}[h!] 
\begin{tabular}{ | c | c | } 
    \hline 
    $\alpha_1$ &  $3.34\times10^5 (T - 381)~\mathrm{m^2 N / C^2 } $ \\
    \hline 
    $\alpha_2$ & $4.69\times10^6 (T - 393) - 2.02\times10^8 ~\mathrm{m^6 N / C^4 } $ \\  
    \hline 
    $\alpha_3$ & $-5.52\times10^7 (T - 393) + 2.76\times10^9 ~\mathrm{m^{10} N / C^6 } $ \\  
    \hline 
    $\kappa$ & $6\times10^{-11} ~\mathrm{m^3 / F } $ \\  
    \hline 
    $\nu $ & $2.5\times10^3 ~\mathrm{m^2 N s / C^2 } $ \\ 
    \hline 
    $\delta x $ & $ 4\times10^{-10 }  ~ \mathrm{m }  $ \cite{structure}  \\ 
    \hline 
    $\delta t $ & $ 1\times10^{-6 }  ~ \mathrm{s }  $ \\ 
    \hline 
\end{tabular} 
\caption{LGD coefficients for barium titanate, taken from Refs.~\cite{LGDcoeffs,structure}. Note that the form of the  thermodynamic potential $V[P]$  depends on the temperature $T$ from the $\alpha_i$ coefficients. The shapes of the potential are shown in Fig.~\ref{fig:potentialshapes}(a). \label{tb:coefficients}}
\end{table}

We now assume that the dynamics of $P$ are not conserved and that the polarization is free to flip in the material due to thermal fluctuations or the applied field $E$. In this case, the dynamics associated with the approach to thermal equilibrium with respect to the free energy density $\psi \equiv \psi[P]$ reads
\begin{equation}
\partial_t P   =- \frac{1}{\nu}\, \frac{\delta \psi }{\delta P} +N \sqrt{\frac{2k_BT }{\nu}}\,\eta(\mathbf{x},t), \label{eq:Langevin}
\end{equation}
where $\psi$ is given in Eq.~\eqref{eq:LGDpotential}, $\nu $ is the ``viscosity'' associated with relaxations of $P$,\ $N$ is a dimensionless noise magnitude,  and $\eta(\mathbf{x},t)$ is a Gaussian white noise with correlations
\begin{equation}
\langle{\eta(\mathbf{x},t)\eta(\mathbf{x}',t')}\rangle = \delta(\mathbf{x}-\mathbf{x}')\delta(t-t').
\end{equation}
 The temperature energy scale $k_BT$ appears here due to the fluctuation-dissipation theorem for equilibrium thermal noise  \cite{tauberbook}, so that a pure thermal noise corresponds to $N=1$.  We will be able to test the effects of varying spatiotemporal noise by varying $N$.  In experiments, such a variable noise may come from an external perturbation such as electrical noise in the probing device used to measure the position of the domain wall. 

 Eq.~\eqref{eq:Langevin} is the equation of primary interest in this study. We will consider the equation in two dimensions, which would correspond to the dynamics of a very thin ferroelectric film.  An important point here is that  Eq.~\eqref{eq:Langevin} is not well-posed in $d \geq 2$ dimensions and requires an appropriate discretization \cite{wellposed}. The discretization introduces a microscopic lengthscale $\delta x$ at which we expect our coarse-grained, phenomenological theory to break down. This should be of order the lattice spacing of the material, which sets $\delta x = 0.4~$nm.

Spatial disorder may be included in Eq.~\eqref{eq:Langevin} in a variety of ways by choosing some of the coefficients in our potential $V[P]$ [or  the gradient coefficient $\kappa$ in Eq.~\eqref{eq:LGDpotential}] to have some random spatial variation. We choose a model of disorder in which the potential $V[P]$ has a spatially-varying magnitude, as was considered for ferromagnetic materials in Ref.~\cite{noisymagneticwalls}. We will describe this approach in more detail in the next section, where we go over in more detail both the   discretization scheme and a convenient coordinate transformation that allows us to analyze moving domain walls.

\subsection{Co-moving frame and discretization}

We will primarily be interested in domain walls moving under the influence of an electric field $E$. This presents some numerical challenges, as our numerical solution necessarily has a finite domain, and the moving domain wall will not remain within the computational domain. Usual solutions, such as introducing periodic boundary conditions, do not work because such conditions are not compatible with a single domain wall (since it connects regions of $P=+P_0$ with $P=-P_0$). To address this challenge, we construct initial conditions such that our domain walls run along the $y$-axis, as shown in Fig.~\ref{fig:domainwall}(c).   Then, the applied field $E$ will generate domain wall motion along the $x$ axis, so it is convenient to  transform variables from $(x,y,t)$ to $(u,y,t)$,    with $u = x-vt$. If the coefficient  $v$ is chosen to be the domain wall speed, then the domain wall will remain stationary in this frame.  This allows us to use fixed boundary conditions ($P \rightarrow \pm P_0$) along the $x$ direction and periodic boundary conditions along the $y$ direction.

For concreteness, we consider a thin film so that our dynamical equation reduces to two spatial dimensions $(x,y)$. Transforming $x$ to $u=x-vt$ yields, after combining    Eq.~\eqref{eq:Langevin} with  Eq.~\eqref{eq:LGDpotential} for the potential $\psi$, a stochastic evolution equation for $P \equiv P(u,y,t)$, the local polarization in the frame moving with speed $v$:
\begin{eqnarray}
\partial_t P & = & D( \partial^2_yP+\partial^2_uP)+v \partial_u P-\nu^{-1} [1+\chi \xi(u,y)]V'(P) \nonumber \\ & &{} +\frac{E}{\nu}+N \sqrt{\frac{2k_BT }{\nu}}\,\eta(u,y,t), \label{eq:Langevin2}
\end{eqnarray}
 where $D=\kappa/\nu \approx 4.8 \times 10^{-14 }~\mathrm{m}^2/\mathrm{s}$ is a diffusion coefficient and $\xi(u,y)$ is a quenched spatial disorder, multiplied by a dimensionless magnitude $\chi$ which we will vary in the following. The spatiotemporal (thermal) noise $\eta $ is evaluated as usual with correlations $\langle\eta(u,y,t)\eta(u',y',t') \rangle=\delta(y-y')\delta( u-u' )\delta(t-t')$. The spatial noise will be specified explicitly below, where we discuss the discretization of the stochastic differential equation in Eq.~\eqref{eq:Langevin2}, for which we will implement  a Euler-Maruyama finite difference integration scheme.

The thin film is discretized into a lattice of sites with spacing given by $\delta x =0.4~\mathrm{nm}$, corresponding approximately to the lattice constant of barium titinate. Then, the spatial derivatives in Eq.~\eqref{eq:Langevin2} may be estimated by finite differences on the square lattice. The local polarization, $P$, will have  values $P^{(t)}_{i,j}$ at each lattice site $(i,j)$ and at each time step $t$. The differential equation, Eq.~\eqref{eq:Langevin2}, then, is approximated on the lattice via the following finite-difference scheme: The polarization $P^{(t+1)}_{i,j}$ at the next time step, $t+1$, is generated via 
\begin{widetext} 
\begin{equation} 
\begin{split} 
    P_{i,j}^{(t+1)} & = P_{i,j}^{(t)}  + \delta t\, \bigg\{ \frac{D}{(\delta x)^2}(P_{i+1,j}^{(t)}+P_{i-1,j}^{(t)}+P_{i,j+1}^{(t)}+P_{i,j-1}^{(t)}- 4 P_{i,j}^{(t)})+\frac{ v}{\delta x}(P_{i,j}^{(t)} - P_{i-1,j}^{(t)})\\ 
    & \quad {}  - \nu^{-1}(1+ \chi \xi_{i,j}) \big(2 \alpha_1 P_{i,j}^{(t)}+ 4 \alpha_2 [P_{i,j}^{(t)}]^3  + 6 \alpha_3 [P_{i,j}^{(t)}]^5\big)  +  \frac{E  }{\nu}+ N \sqrt{\frac{2k_{B}T  }{\Delta t (\delta x)^3 \nu}}\,\eta_{i,j}^{(t)}  \bigg\},       
\end{split}  \label{eq:discrete}
\end{equation} 
\end{widetext} 
 where $\eta_{i,j}^{(t)}$ are independently-sampled, Gaussian random variables at each lattice site with unit variance, which follows directly from the correlations for $\eta$ in Eq.~\eqref{eq:Langevin2}. Then, for the spatially quenched disorder, we choose $\xi_{i,j}$ to be independent, uniform random variables between $-1$ and $1$, which remain the same values throughout the time evolution.  This choice corresponds to a local variation in the magnitude of the thermodynamic potential $V[P]$. A similar approach was used to model ferromagnetic systems \cite{noisymagneticwalls,LGwalls}, where this noise was shown to pin domain walls.  The coefficient $\chi$ sets the noise magnitude.  The velocity $v$ is tuned until a steady-state is achieved and the domain wall remains at the center of the simulation box (for a particular fixed value of $E$). 

\begin{figure}[!ht] 
\includegraphics[width=3.3in]{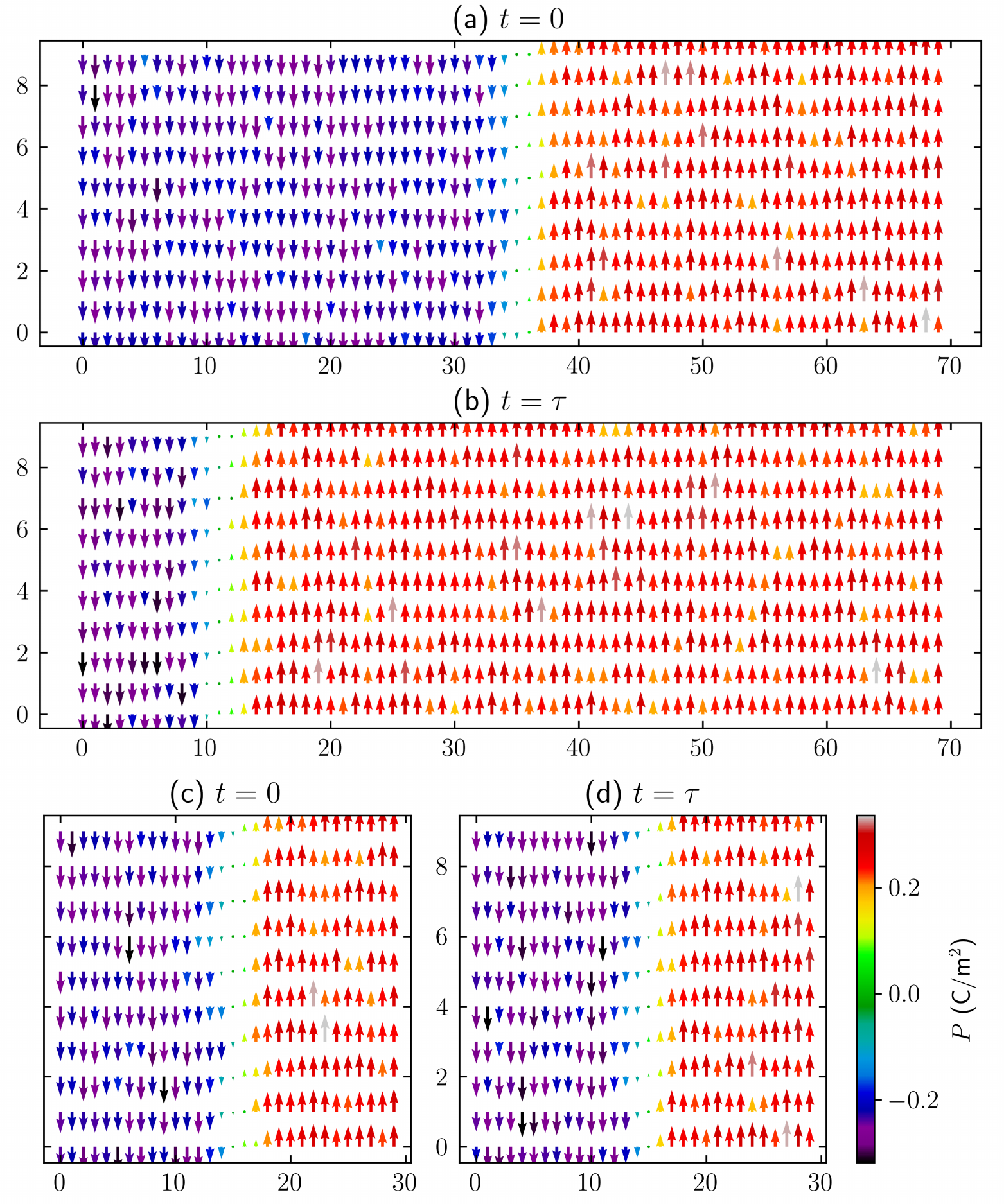} 
\caption{\label{fig:comoving} We show successive snapshots of a simulated wall (in the static `lab' frame) at an initial time $t=0$ (a) and a final time $t=\tau$ (b) after $75000 $ timesteps under an applied field $E=1050~\mathrm{V}/\mathrm{cm}$. Note that the domain wall will leave the simulation window at later tames. Using the comoving frame, shown in (c) and (d), it is possible to simulate a moving domain wall over a much longer period of time by adjusting the  velocity $v$ [see Eqs.~(\ref{eq:Langevin2},\ref{eq:discrete})]. The distances are measured in lattice spacings (taken to be 0.4 nm).   } 
\end{figure}

Our computational domain is an $L_x \times L_y$ square lattice with $L_x=70$ and $L_y=10$ sites, unless otherwise stated. Our initial condition is a single domain wall along the $y$-axis in the middle of the computational domain, as shown in the light green dashed lines in Fig.~\ref{fig:domainwall}(c-e). The boundaries are fixed in the (horizontal) $x$ direction and periodic in the (vertical) $y$ direction:  We set $P=-P_0$ on the left and $P=P_0$ on the right, using the calculated form of $P_0$ in Eq.~\eqref{eq:P0}.  Plots of simulated domain walls are shown in Fig.~\ref{fig:domainwall}(c-e) and in Fig.~\ref{fig:comoving}.

Further, the domain wall velocity is measured in the Galilean boosted frame where the boost velocity is dynamically determined to match domain wall movement. This is done by periodically evaluating the domain wall position and adjusting the velocity to take into account any displacement observed over the observation time window. Eventually, the parameter $v$ in Eq.~\eqref{eq:discrete} settles to a steady state value (with some fluctuations) which we can report as the domain wall speed.    Fig.~\ref{fig:comoving} shows simulations of the system with thermal noise in both the static (a,b) and boosted (c,d) frames.

 \section{Results}
 
 \subsection{Zero noise limit}
 
 It is instructive to look for domain wall solutions to Eq.~\eqref{eq:Langevin} in the absence of noise ($\eta=0$).   We find that the equation, for small applied fields $E$, admits stationary solutions describing \textit{domain walls} between regions of $P=P_0$ and $P=-P_0$. These domain wall solutions  can be calculated for arbitrary values of the coefficients in Eq.~\eqref{eq:potential} below the phase transition temperature $T_c$. The solution method is  discussed in more detail in the Appendix. These solutions have been studied for some time, with work dating back decades to the analysis of domain walls in shape-memory alloys \cite{Falk1983}, to more recent, general analysis of periodic arrays of such domain wall solutions \cite{phi6kinks,phi6kinks2}.

As detailed in the Appendix, we find that the stationary shape is
\begin{equation}
P(x)= \frac{P_0  \sqrt{1-\gamma  }\,\tanh \left(\frac{x}{\xi}\right)}{  \sqrt{1-\gamma \tanh ^2\left(\frac{x}{\xi}\right)}}, \label{eq:profilezeroT}
\end{equation}
where $\gamma=   \left[|\alpha_2|(\alpha_2^2- 3 \alpha_1 \alpha_3)^{-1/2}+1\right]/3$ and $\xi$ is a measure of the domain wall width, given by 
\begin{equation}
\xi=\frac{  \sqrt{\kappa \alpha_3}\,(3 \gamma-1)}{ |\alpha_2| \sqrt{  2\gamma} \, }.
\end{equation}
At $T=300~\mathrm{K}$, we have $\gamma\approx 0.541$ and $\xi \approx 0.646~\mathrm{nm}$.  In a second-order phase transition, we would expect $\xi$ to diverge at $T=T_c$ (so that $\gamma \rightarrow 1$). Instead, the  domain wall width approaches a fixed $\xi = \sqrt{\kappa \alpha_3}/|\alpha_2|$ and the function $P(x)$ develops a large plateau at $P=0$.  The size $\ell$  of the plateau can be estimated from Eq.~\eqref{eq:profilezeroT}:
\begin{equation}
\ell \approx \xi \tanh^{-1}[\sqrt{3 \gamma-2}].
\end{equation}
In other words, near the phase transition, we expect our domain walls to develop paraelectric regions which diverge in size as we approach the transition.  This is  sensible as the domain wall is a natural place for the paraelectric phase (which become favorable for $T>T_c$) to nucleate and grow.  At exactly $T=T_c$, the potential has three wells corresponding to three equally favorable phases ($P=0,\pm P_0$). In this case, the noiseless LDG equation admits stable half-wall (often also called half-kink) solutions which may propagate independently \cite{phi6kinks2}.  We now will analyze the effects fluctuations near such a transition.

\subsection{Noisy domain wall results}

\begin{figure}[!ht]
\includegraphics[width=3.3in]{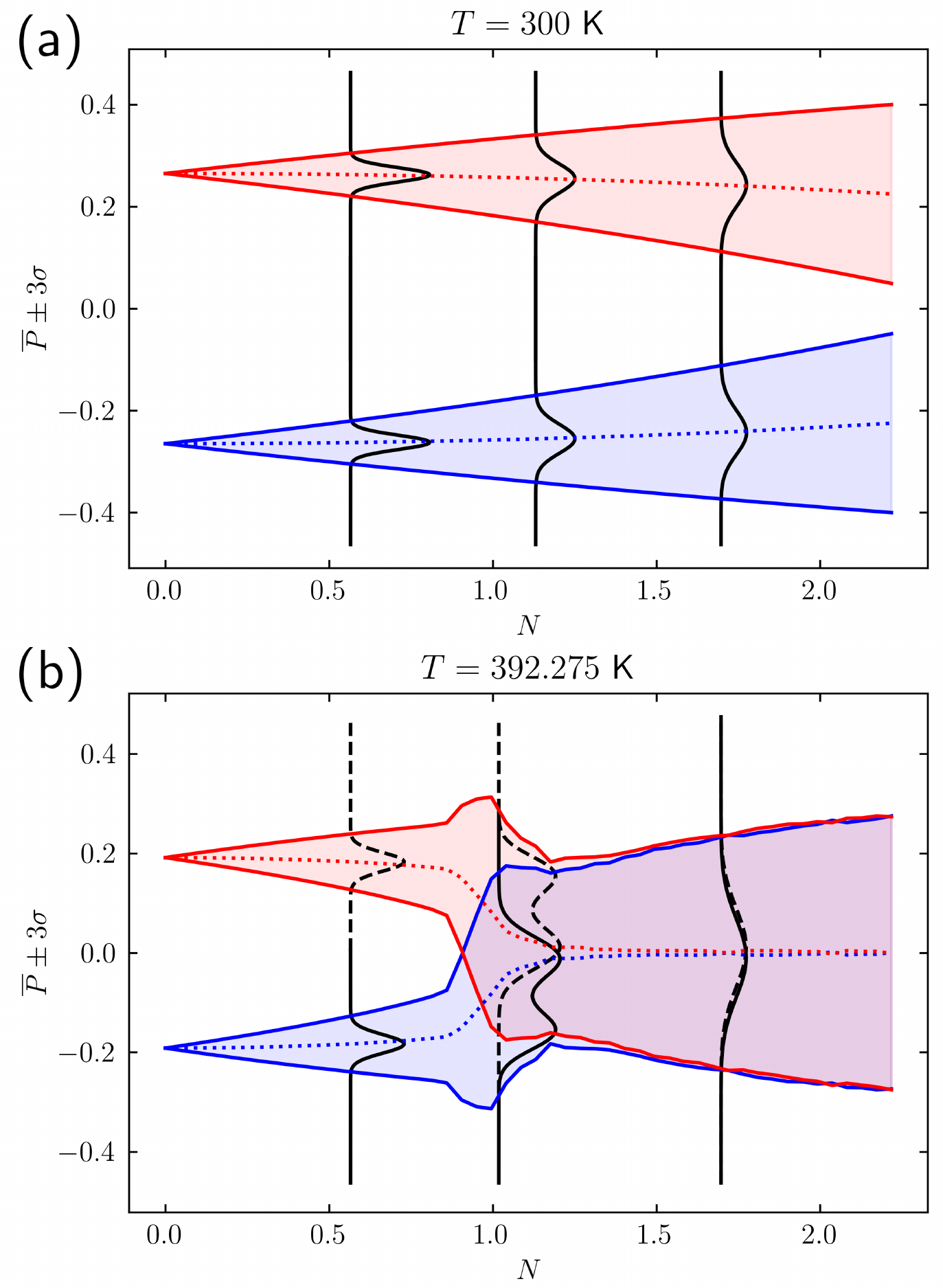}  
\caption{\label{fig:Pdistributions} Polarization distribution of a region of either positive (red) or negative (blue) polarization evolved over a fixed system size and time period with thermal noise given by noise power $N$. The mean polarization plus or minus 3 standard deviations is plotted as the red/blue lines with shading in between, and the black lines show histograms of the polarization at certain $N$ values. (a) gives the distributions at 300~K, and Figure (b) gives the distributions at just below the critical temperature: $T=392.275~\mathrm{K} < T_c = 392.277\ldots~\mathrm{K}$ . The system size is 500$\times $ 500 lattice spacings and polarization distributions are taken from regions of size 200$\times $ 500 lattice spacings in both the positive and negative polarization regions. The distributions are taken over a 500 timestep period after evolving the system for an initial 10000 timesteps.  } 
\end{figure}

When considering the effect of noise on the domain wall, we first focus on thermal noise. The thermal noise magnitude is controlled by the constant $N$ in   Eq.~\eqref{eq:discrete}. Thermal noise tends to broaden the distribution of polarizations in the system. This is shown in Fig.~\ref{fig:Pdistributions}, where the polarization distributions are shown in regions of primarily $P=P_0$ (red distribution) and $P=-P_0$ (blue distribution). Note that as we increase the noise power $N$, the distributions widen. At $T$ just below $T_c$, as shown in Fig.~\ref{fig:Pdistributions}(b),  the distributions begin to overlap and develop a peak at $P=0$ when $N \gtrsim 1$. This means that increasing the noise can lead to a transition and the loss of the ferroelectric state.
For the material considered here, unsurprisingly, the noise magnitude that corresponds to thermal fluctuations $(N=1)$ is just at the cusp of the transition.

We first observe the behavior of the polarization far from $T_c$. In Fig.~\ref{fig:Pdistributions}(a), the mean polarization $\pm 3$ standard deviations is plotted as a function of noise power $N$ (red and blue regions), as well as histograms of the polarization at certain noise power values (black solid lines). Far from $T_c$, the thermal noise is unlikely to cause spontaneous switching, resulting in the linear increase of the distribution width.  However, near $T_c$, when the potential barrier height is on the order of the thermal noise, spontaneous switching can occur, corresponding to a shift in the mean polarization and a spike in the standard deviation. At this point, shown in Fig.~\ref{fig:Pdistributions}(b), the histograms of the polarization show two peaks in both regions of polarization (positive and negative): one at the potential minimum near the corresponding polarization $P=\pm 0.19$, and one at $P=0$, the metastable paraelectric phase. The two distributions (e.g., to the left and right of a domain wall) are shown by the solid and dashed black lines in Fig.~\ref{fig:Pdistributions}(b).  In other words, for sufficiently large magnitude, $N \geq  1$, the thermal fluctuations will generate a transition between the potential well minimum at $P = \pm P_0$ and the one at $P=0$. For $N \gtrsim  1.5 $, the positive and negative polarization regions become indistinguishable as the thermal noise makes the potential barrier negligible, so there is a single peak in the polarization distribution near the central minimum at $P=0$.

\begin{figure}[!ht]
\includegraphics[width=3.2in]{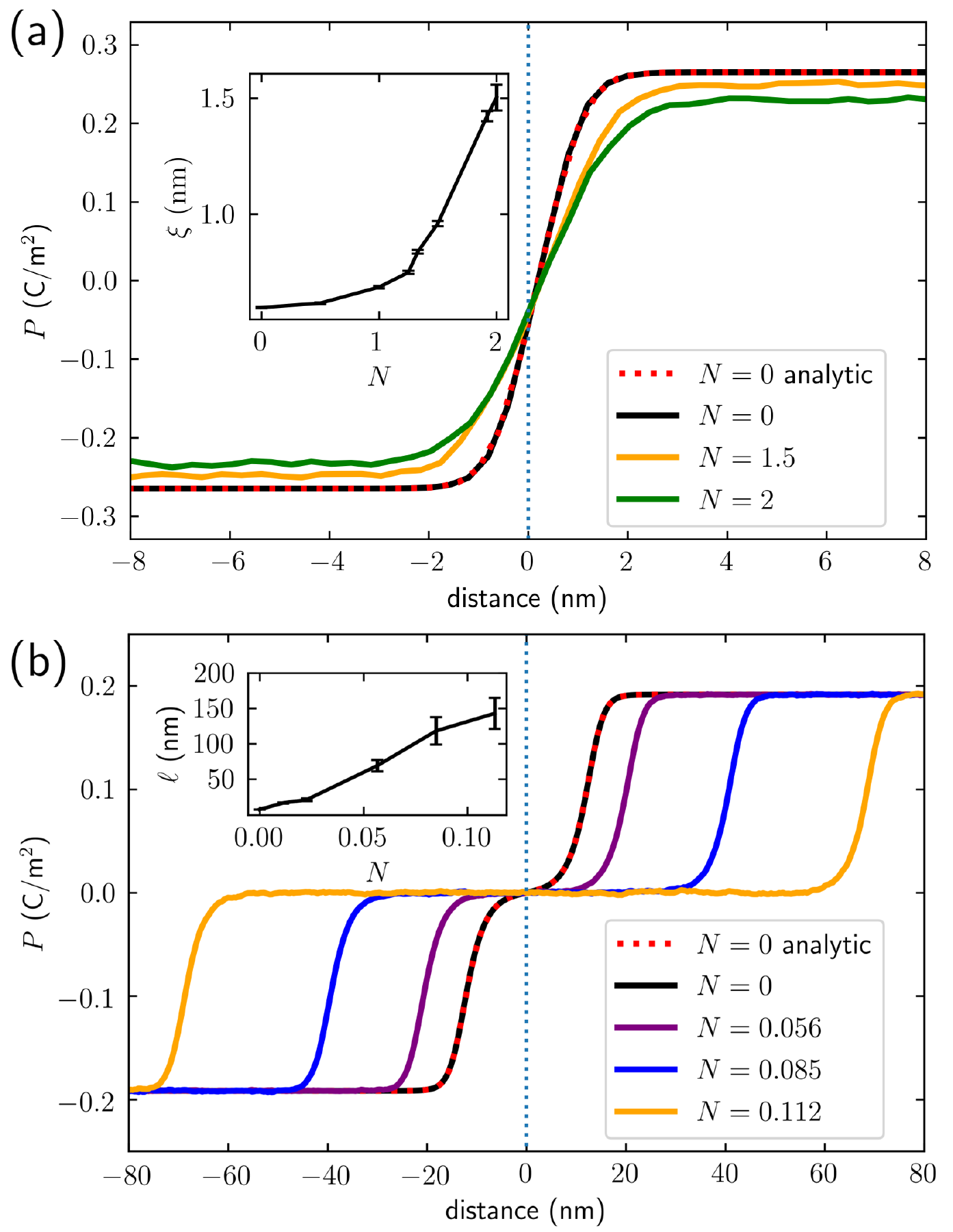} 
\caption{\label{fig:domainwallshapes} (a) Simulated domain wall shapes for various values of the thermal noise magnitude $N$ at a fixed temperature $T=300$~K.  Note that the $N=0$ form is known analytically, and matches our numerical solution. (b) Simulated domain wall shapes for different noise magnitudes for domain walls just below the critical temperature $T=392.275~\mathrm{K} < T_c=392.277\ldots~\mathrm{K}$. } 
\end{figure}

The effect of thermal noise is also observed in the shape of the $180^{\circ}$ domain wall, shown in Fig.~\ref{fig:domainwallshapes}. Here, the domain wall is simulated without an applied electric field and is allowed to evolve until it relaxes to a constant shape. Figure~\ref{fig:domainwallshapes} gives the shape of the domain wall as a function of thermal noise for temperature $T$ well below the transition temperature $T_c$ in (a) and for $T$ \textit{just} below $T_c$ in (b). In the latter case, for $N \lesssim 0.87 $, the size of the domain wall increases with increasing $N$, but for $N \gtrsim 0.87$, the domain wall dramatically broadens and effectively splits into two separate, half-domain walls, as the thermal noise is large enough to cause spontaneous switching to the $P=0$ minimum.   The $P=0$ region of the domain wall can grow to span nearly the full simulation box, as shown in Fig.~\ref{fig:domainwallshapes}(b) and the inset. By contrast, when the potential consists of just two deep wells at $P=\pm P_0$, the thermal noise more modestly (but significantly) increases the domain wall width, as shown in Fig.~\ref{fig:domainwallshapes}(a) and the inset.

\begin{figure}[!ht]
\includegraphics[width=3.3in]{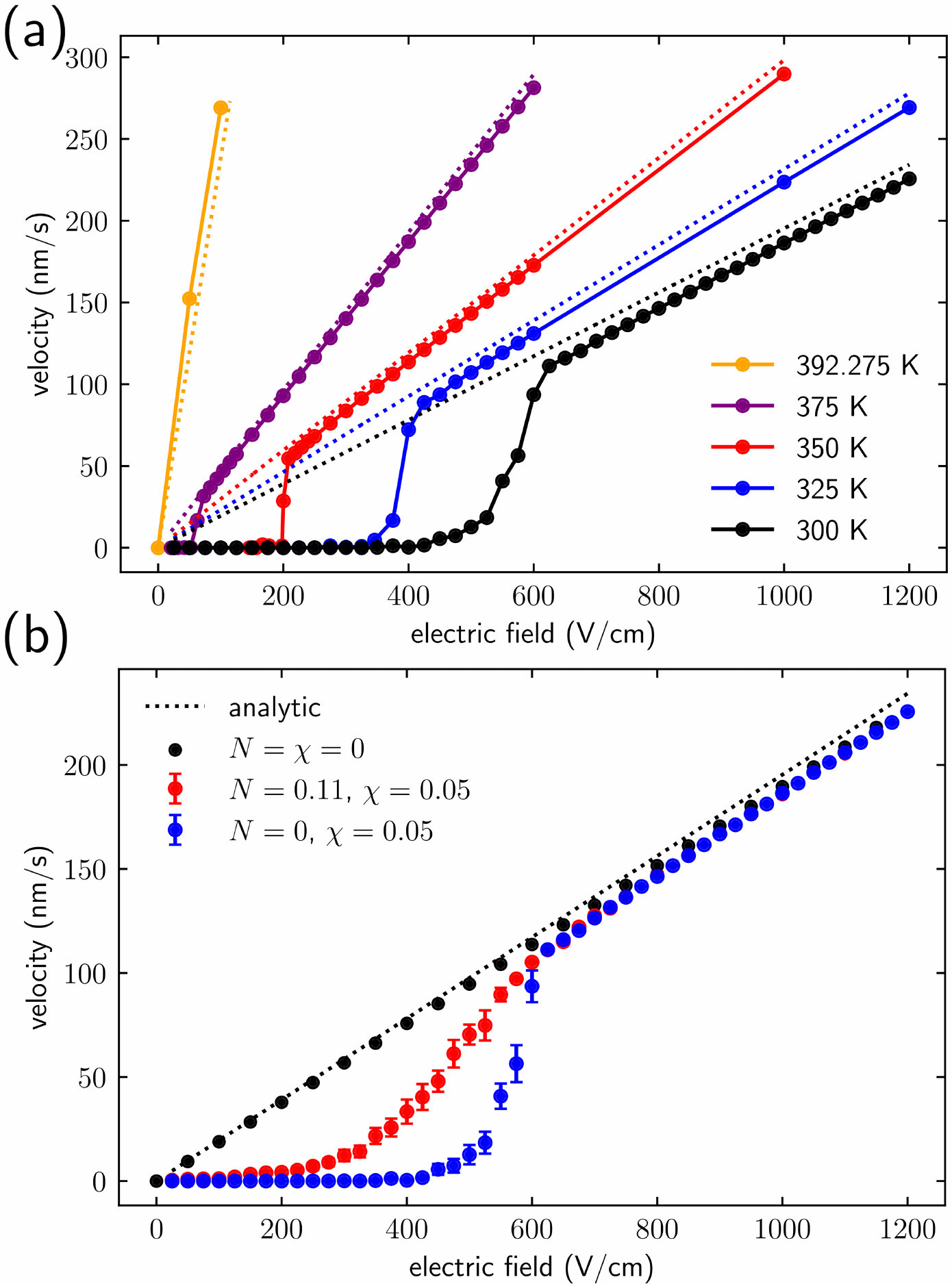} 
\caption{\label{fig:velocities} (a) Domain wall velocity as a function of electric field for systems with spatial noise $\chi = 0.05$ for various temperatures approaching $T_c$ (without thermal noise). The dotted lines show the analytic result for the domain wall velocity for each temperature (for $\chi=0$). (b) Domain wall velocity for the $T=300$~K case for various noise magnitudes indicated in the legend. Bars denote $1$ standard error of the fluctuating velocity as measured in the comoving frame. The system size is $50 \times 50 $ lattice spacings and the motion was measured over $1.5 $ million timesteps. The dotted line is the analytic result for the noiseless domain wall speed.} 
\end{figure}

Now we consider the effect of only spatial noise on the system. The results are shown in Fig.~\ref{fig:velocities}(a).  When we are far below the transition temperature $T_c$ and have a double-well potential, there is a pinning effect due to the spatial noise [darker shaded curves in Fig.~\ref{fig:velocities}(a)].  When the domain wall velocity is observed as a function of applied electric field, the spatial noise prohibits the domain wall from moving until some threshold electric field $E_{dp}$ is reached. However, as the temperature increases towards $T_c$ for constant spatial noise $\chi = 0.05$ [yellow curve in Fig.~\ref{fig:velocities}(a)], the threshold electric field decreases significantly. This effect and the comparison to analytical predictions for velocity from the previous section are shown in Fig.~\ref{fig:velocities}(a).  Note that the domain wall velocity at $T=T_c$ does not diverge as near a second order phase transition point. Instead, the domain wall decouples into two, independently-moving half walls between $P=\pm P_0$ and $P=0$ regions. The limiting value of the velocity at $T=T_c$ is approximately $12$ times greater than the value at $300$~K, where the potential $V[P]$ has the form of a deep double well pcompare black and yellow curves in Fig.~\ref{fig:velocities}(a)]b.

Upon depinning, the velocity quickly approaches the analytical result [dotted lines in Fig.~\ref{fig:velocities}], such that the spatial noise is no longer relevant for electric fields $E>E_{dp}$. As the temperature is increased to the critical point $T_c$, the depinning field $E_{dp}$ decreases by many orders of magnitude, as the potential barrier between the two polarization states $\pm P_0$ becomes quite small.  Thermal noise also serves to depin the domain wall, as shown in Fig.~\ref{fig:velocities}(b) for the $T=300~\mathrm{K}$ case. Note how for values of $E<E_{dp} \approx 580~\mathrm{V}/\mathrm{cm}$, the domain wall with just spatial quenched disorder (blue points) has the smallest domain wall velocities, jumping to nearly zero below the depinning threshold. Adding some thermal noise [$N>0$ for red points in Fig.~\ref{fig:velocities}(b)] can lead to a many-fold increase in the velocities below the depinning threshold. Finally, note that all the noises are irrelevant for large fields $E>E_{dp}$.

The results shown in Fig.~\ref{fig:velocities}(b) recapitulate the known behavior of domain walls in noisy double-well potentials \cite{noisymagneticwalls,LGwalls,creepreview}. While these results are important for understanding ferroelectric switching below the critical temperature, it is clear that the situation is quite different near $T_c$ where the metastable paraelectric phase plays a key role. To understand these differences, we consider the sextic potential Eq.~\eqref{eq:potential} just below the critical temperature at $T=392.275 $ and a double well potential $V_4[P]$ with equal potential barriers and curvatures at the bottom of each well:
\begin{equation}
V_4(P) =-\frac{\beta^2}{2\gamma} P^2+\frac{27 \beta }{8}P^4+\frac{\beta ^3}{54 \gamma ^2 } 
\label{eq:DWTW}
\end{equation}
where we choose $\beta=2.05 \times 10^8 ~\mathrm{m^6 N / C^4 } $ and $\gamma=2.80 \times 10^9 ~\mathrm{m^{10 } N / C^6 } $. In other words, we may consider a ``double well'' (DW) version of the triple well potential that occurs at $T$ just below $T_c$.   We may now set $V(P)=V_{4}(P)$ in Eq.~\eqref{eq:Langevin2}  and compare the results to the triple well (TW).

\begin{figure}[!ht]
\includegraphics[width=3.3in]{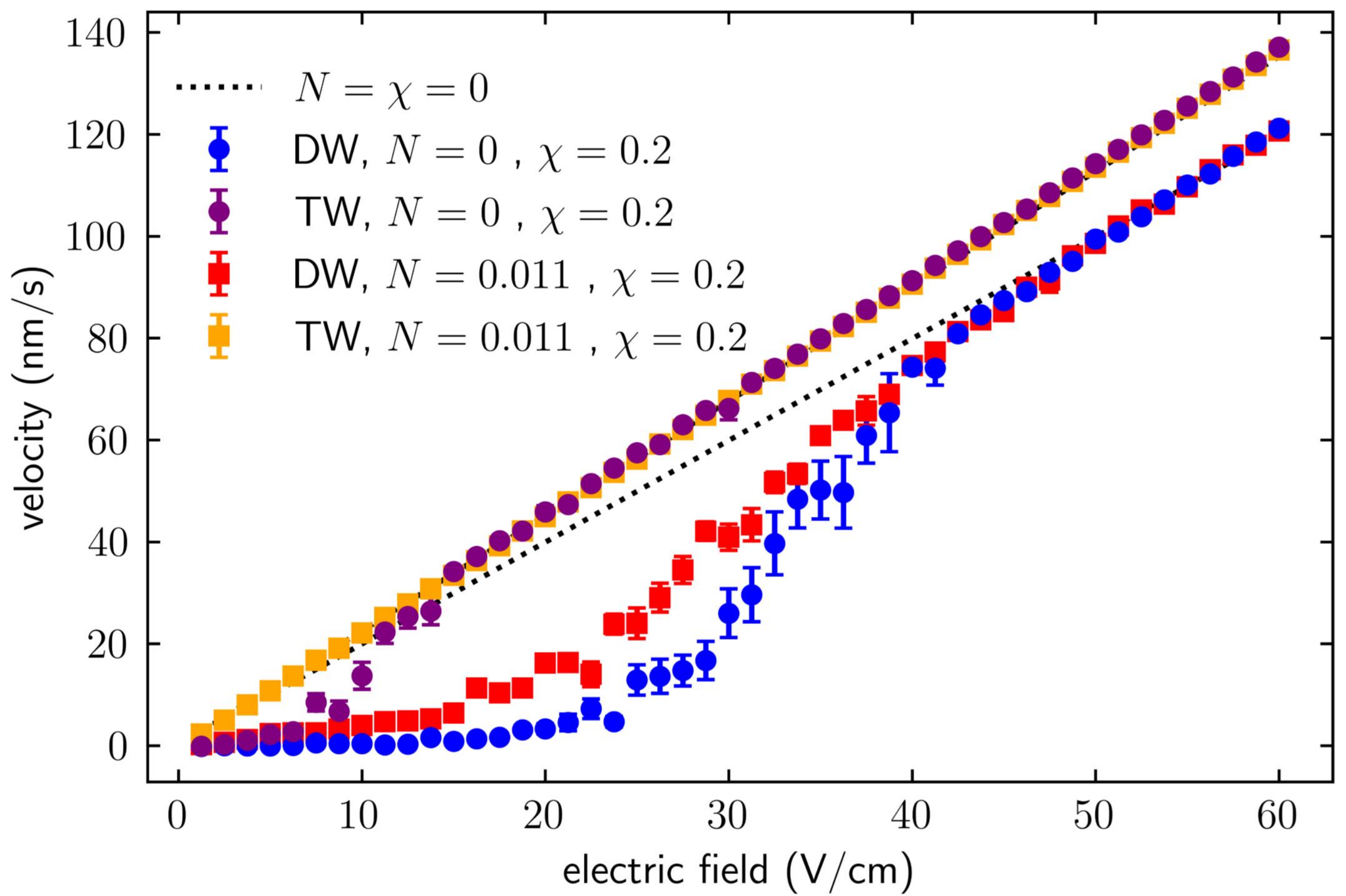} 
\caption{\label{fig:DWTW} Domain wall speed as a function of applied field $E$ for both the double well (DW) and triple well (TW) potentials [see Eq.~\eqref{eq:DWTW}], with a spatial quenched noise with magnitude $\chi=0.2$ [see Eq.~\eqref{eq:Langevin2}]. Bars denote the standard error  of the  velocity measurement in the comoving frame. Note that for $N=0$ (circles), both walls are pinned at small fields $E$, with the TW domain wall depinning at a much smaller $E_{dp}$.  When we add some thermal noise with magnitude $N=0.0112$ (squares), the thermal noise can completely depin the TW domain wall, while the DW domain wall remains pinned at small applied fields $E$.  The system size is $50 \times 500 $ lattice spacings and the motion was measured over $6 \times 10^6$ timesteps. } 
\end{figure}

The results are shown in Fig.~\ref{fig:DWTW}. 
First, the domain wall velocity is plotted as a function of applied electric field $E$ for the two potentials (DW and TW) with quenched spatial noise and no thermal noise $N=0$ (circles). The black dotted lines show the domain wall velocities of the noise-free cases. Note that the velocity in the TW case is slightly larger than for the double well (DW). Domain walls in both potentials have a pinning effect for electric fields $E$ below a critical depinning field $E_{dp}$. For larger fields $E>E_{dp}$,  the domain walls depin and thier velocity quickly approaches the noise-free case. The double well (blue circles) has a depinning field nearly 4 times greater than the triple well (purple circles) depinning field, indicating that the metastable minimum in the triple well potential aids in depinning. Next, we can check the ``triple-well-enhanced'' depinning in another way by introducing a thermal noise, which we expect to aid in depinning. We find that with an added thermal noise $N=0.011$ (squares in Fig.~\ref{fig:DWTW}), the domain walls for the TW (yellow squares) completely depin at this level of thermal noise, while the domain walls in the DW potential (red squares) are only partially depined, with the domain wall velocity remaining close to zero below the depinning field.

 \section{Discussion and Conclusion}
 
We have shown that thermal and spatial, quenched noise can strongly influence the domain wall dynamics in a ferroelectric material. We have focused here on a uniaxial ferroelectric and given special attention to the domain wall behavior near the critical temperature $T_c$, where the thermodynamic potential $V[P]$ for  the polarization $P$ exhibits a (meta)stable minimum at $P=0$ (a paraelectric phase). In this case, the domain wall structure is more complicated than what might be expected in a standard double-well potential with minima just at the two stable polarization states $P=\pm P_0$. The low energy modes corresponding to domain wall motion are complicated by the internal structure of the domain wall, with additional low-energy translational modes developing when the half-walls (connecting the $P=\pm P_0$ regions to $P=0$) nearly decouple at $T_c$. The detailed mode structure for these domain wall solutions has been studied previously in the noise-free case in, e.g., Ref.~\cite{phi6kinks}. Here we show that thermal noise may also excite these modes: As shown in Fig.~\ref{fig:domainwallshapes}(b), the domain wall consists of a bound pair of half-walls, which may become well-separated for sufficiently large noise strength $N$ near $T=T_c$.  This precludes the possibility of using a single coordinate (i.e., the center of mass) to describe the domain wall motion, as can be done with the double-well case using an elastic line model \cite{LGwalls}. Our results here elucidate the effects of noise on these more complicated domain wall shapes, which are important not only in the context of ferroelectrics, but in other fields where such models are used. For example, the sextic potential considered here serves as a basic model for a bound pair of quarks, with each half-wall representing a quark \cite{ChristLee}.

We have also shown that the metastable paraelectric phase aids in depinning the domain wall. By comparing a double and triple-well potential, we show that the triple well case has a much smaller depinning field $E_{dp}$ for the same potential barrier heights and the same spatial noise magnitude $\chi$ [see Eq.~\eqref{eq:Langevin2}].  Thermal noise also more readily activates the domain wall motion, overcoming the pinning for much smaller noise magnitudes $N$. We have also shown that thermal noise may strongly modify the distributions of polarizations in the material, nucleating the paraelectric phase as shown in the distributions in Fig.~\ref{fig:Pdistributions}(b).  Note that without noise, the differences between the triple and double well scenarios are relatively minimal, at least in terms of the domain wall mobility. Note in Fig.~\ref{fig:DWTW} that the DW velocities are similar for large $E$. However, the value of the depinning field $E_{dp}$ is strongly suppressed by the presence of the (metastable) paraelectric phase. Thus suggests that noise can be a useful ``knob'' for modifying the switching behavior of a ferroelectric material, especially near the ferroelectric phase transition. Here we considered spatial noise due to, e.g., material defects and spatiotemporal noise due to thermal fluctuations. However, noise may be introduced into these systems externally by applying, for example, a spatially varying and/or time-dependent electric field, as is done to induce noise-assisted responses in non-linear systems via the phenomenon of stochastic resonance \cite{stochresreview}.   Our results here demonstrate that tuning the amount of noise can significantly impact the mobility of domain walls within the material.

\acknowledgments

We thank Michael A. Susner for providing CuInP$_2$S$_6$ and Sn$_2$P$_2$S$_6$ materials. This research was supported in part by an appointment to the Oak Ridge National Laboratory Technical and Professional Internship, sponsored by the U.S. Department of Energy and administered by the Oak Ridge Institute for Science and Education. P. M. and S. M. N. acknowledge support from the U.S. Department of Energy, Office of Science, Basic Energy Sciences, Materials Science and Engineering Division.Piezoresponse force microscopy was conducted at the Center for Nanophase Materials Sciences, (CNMS), which is a US Department of Energy, Office of Science User Facility at Oak Ridge National Laboratory.

\appendix

\section{Domain wall shape and velocity\label{appx:shape}}

In this section we develop the analytic solution, without noise, for the domain wall shape for the $P^6$ potential given by Eq.~\eqref{eq:potential}. The solution can be derived using standard techniques, and has been calculated in a variety of contexts previously \cite{phi6kinks,phi6kinks2,Falk1983}.
We look for long, straight domain walls so that the polarization profile $P$ only depends on a single spatial coordinate $x$.  Let us introduce a dimensionless position $u \equiv \alpha_x x$ and unitless polarization $p \equiv \alpha_p P$. Convenient choices for the scaling factors are, in terms of the parameters in Eq.~\eqref{eq:potential},
\begin{equation}
 \alpha_x=2\left[- \frac{2\alpha_2}{3\alpha_3}\right]^{5/4} \sqrt{\frac{3\alpha_3}{2}} \mbox{ and } \alpha_p= \sqrt{-\frac{3 \alpha_3}{2\alpha_2}}.
\end{equation}
Our equation of motion for the polarization $P \equiv P(x,t)$, Eq.~\eqref{eq:Langevin} (without the noise terms), then becomes, in the presence of an applied field $E$:
\begin{equation}
\frac{\nu}{\alpha_p\alpha_x^2}\frac{\partial p}{\partial t}=g \frac{\partial^2p}{\partial u^2}-\bar{\alpha} p + p^3-p^5 +\frac{E}{\alpha_x^2}, \label{eq:scaledLangevin}
\end{equation}
 where  $g=D \sqrt{-\frac{2\alpha_2}{3 \alpha_3}}$ and  $\bar{\alpha}=\frac{3 \alpha_1 \alpha_3}{4 \alpha_2^2}$. For a single domain wall, we would have the boundary conditions $p=\pm p_0$ for $u \rightarrow \pm \infty$, respectively, with $p_0^2=(1+\sqrt{1-4 \bar{\alpha}})/2$. When $E=0$, we expect to find a stationary domain wall solution with $\partial_t p=0$ in Eq.~\eqref{eq:scaledLangevin}. Integrating Eq.~\eqref{eq:scaledLangevin}  with respect to $u$ yields an equation for the stationary profile $p \equiv p(u)$
\begin{equation}
\frac{g}{2}\left[ \frac{dp}{du} \right]^2=   \frac{\bar{\alpha} p^2}{2} - \frac{p^4}{4}+\frac{p^6}{6}-\left[\frac{\bar{\alpha} p_0^2}{2}  - \frac{p_0^4}{4}+\frac{p_0^6}{6}\right], \label{eq:SSprofile}
\end{equation}
where we have stipulated that  $dp/du \rightarrow 0$ for $u \rightarrow \pm \infty$. This equation can be integrated again and a stationary shape for the domain wall derived:
\begin{equation}
p(u)= \pm \frac{\sqrt{ \tilde{\alpha}\left[ \sqrt{1-4 \bar{\alpha} }-\frac{1}{2}\right]}  \tanh \left(\frac{ \tilde{\alpha}(1-4\bar{\alpha}) \,u}{2 \sqrt{g}}\right)}{  \sqrt{3-  \tilde{\alpha} \tanh ^2\left(\frac{\tilde{\alpha}(1-4\bar{\alpha})\,u}{2 \sqrt{g}}\right)}}, \label{eq:scaledDWshape}
\end{equation}
where $\tilde{\alpha}=1+(1-4\bar{\alpha})^{-1/2}$. This form reduces to the expression given by Eq.~\eqref{eq:profilezeroT} in the main text. There is a characteristic size of the domain wall here, given in terms of the original LDG parameters as
\begin{equation}
\xi =\frac{  \sqrt{\kappa}}{ \left[- \frac{2\alpha_2}{3\alpha_3}\right]  \sqrt{\frac{3\alpha_3}{2}}\sqrt{1- \frac{3 \alpha_1 \alpha_3}{  \alpha_2^2} +\sqrt{1-\frac{3 \alpha_1 \alpha_3}{  \alpha_2^2} }} \, }.
\end{equation}
 Note that our simulations will require that our lattice spacing $\delta x$ is  smaller than $\xi$, so that the domain wall shape may be fully resolved. This is  satisfied by our choice $\delta x=0.4~\mathrm{nm}$.

Let us now calculate the speed $v$ for the domain wall without noise.  We look for solutions to Eq.~\eqref{eq:scaledLangevin} (with a constant term corresponding to the applied electric field $E$) where $p(u,t)$ can be replaced by a function (the stationary wall profile, in particular) of a single variable $p[X = u-u_0(t)]$, where $u_0(t)$ would represent the center of the soliton, which will move as a function of time. Substituting such an ansatz into Eq.~\eqref{eq:scaledLangevin} (without noise) yields
\begin{equation}
-\frac{\nu \dot{u}_0}{\alpha_p\alpha_x^2}\frac{dp}{dX}=g \frac{d^2p}{dX^2}-\bar{\alpha} p + p^3-p^5 +\frac{E}{\alpha_x^2} , \label{eq:scaledLangevin2}
\end{equation}
where $\dot{u}_0$ is the velocity of the domain wall in the rescaled coordinates. If the applied electric field $E$ is small, then the shape of the domain wall should remain close to the stationary domain wall profile given by   Eq.~\eqref{eq:scaledDWshape}.  Substituting the stationary profile for $p(X)$ into   Eq.~\eqref{eq:scaledLangevin2} 
yields an equation for the domain wall velocity:
\begin{equation}
\dot{u}_0=- \frac{2p_0\alpha_pE}{\nu}\left[\int_{-\infty}^{\infty}\left( \frac{dp}{dX} \right)^2\,\mathrm{d}X \right]^{-1} \equiv \frac{2 \Xi(\bar{\alpha})}{\sqrt{g}} , \label{eq:scaledvelocity}
\end{equation}
where we have introduced a convenient function $\Xi(\bar{\alpha})$ which may be evaluated by performing the integration in Eq.~\eqref{eq:scaledvelocity}  using the expression for $p(X)$ found in  Eq.~\eqref{eq:scaledDWshape}. 
Substituting in for the various coefficients, we find the velocity $v$ in terms of the parameters given in the main text:
\begin{equation}
v= \frac{  3\alpha_3 \sqrt{ D}}{ \nu(-2\alpha_2)^{3/2} }   \frac{ \sqrt{ 1+\sqrt{1- \frac{3 \alpha_1 \alpha_3}{  \alpha_2^2}} }  }{ \Xi(\frac{3 \alpha_1 \alpha_3}{4 \alpha_2^2})} \, E \label{eq:analyticvel}
\end{equation}
We may also evaluate this velocity at the critical temperature $T=T_c$, where $\bar{\alpha}=\bar{\alpha}_c=3/16$. In this case, Eq.~\eqref{eq:analyticvel} reduces to 
\begin{equation}
v(T=T_c)=\frac{    8\alpha_3 \sqrt{ D}}{ \nu(- \alpha_2)^{3 /2} } \, E.   
\end{equation}
Therefore,  we find a constant velocity at the phase transition given by $v \approx 7~\mu\mathrm{m}/\mathrm{s}$ for $E = 10^5~\mathrm{V}/\mathrm{cm}$.

\bibliographystyle{apsrev}
\bibliography{DWMBib}

\end{document}